\documentclass[12pt,cls,onecolumn]{IEEEtran}
\usepackage{subfigure,cite,graphicx,psfig,amsmath,amssymb,mathrsfs,epsf}

\begin{document}

\title{Parallel Opportunistic Routing in Wireless Networks}
\author{\large Won-Yong Shin, \emph{Member}, \emph{IEEE}, Sae-Young Chung, \emph{Senior Member}, \emph{IEEE}, \\and Yong H.
Lee, \emph{Senior Member}, \emph{IEEE} \\
\thanks{The material in this paper was presented in part at the IEEE
International Symposium on Information Theory, Nice, France, June
2007.}
\thanks{W.-Y. Shin was with the School of EECS, KAIST, Daejeon 305-701, Republic of Korea. He is now with the School of Engineering and Applied Sciences, Harvard
University, Cambridge, MA 02138 USA
(E-mail:wyshin@seas.harvard.edu).}
\thanks{S.-Y. Chung and Y. H. Lee are with the School of EECS, KAIST, Daejeon 305-701, Republic of Korea
(E-mail: sychung@ee.kaist.ac.kr; yohlee@ee.kaist.ac.kr).}
        } \maketitle


\markboth{Under Review for Possible Publication in IEEE Transactions
on Information Theory} {Shin {\em et al.}: Parallel Opportunistic
Routing in Wireless Networks}


\newtheorem{definition}{Definition}
\newtheorem{theorem}{Theorem}
\newtheorem{lemma}{Lemma}
\newtheorem{example}{Example}
\newtheorem{corollary}{Corollary}
\newtheorem{proposition}{Proposition}
\newtheorem{conjecture}{Conjecture}

\def \diag{\operatornamewithlimits{diag}}
\def \min{\operatornamewithlimits{min}}
\def \max{\operatornamewithlimits{max}}
\def \log{\operatorname{log}}
\def \max{\operatorname{max}}
\def \rank{\operatorname{rank}}
\def \out{\operatorname{out}}
\def \exp{\operatorname{exp}}
\def \arg{\operatorname{arg}}
\def \E{\operatorname{E}}
\def \tr{\operatorname{tr}}
\def \SNR{\operatorname{SNR}}
\def \dB{\operatorname{dB}}
\def \ln{\operatorname{ln}}

\def \be {\begin{eqnarray}}
\def \ee {\end{eqnarray}}
\def \ben {\begin{eqnarray*}}
\def \een {\end{eqnarray*}}

\begin{abstract}
We study benefits of opportunistic routing in a large wireless ad
hoc network by examining how the power, delay, and total throughput
scale as the number of source--destination pairs increases up to the
operating maximum. Our opportunistic routing is novel in a sense
that it is massively parallel, i.e., it is performed by many nodes
simultaneously to maximize the opportunistic gain while controlling
the inter-user interference. The scaling behavior of conventional
multi-hop transmission that does not employ opportunistic routing is
also examined for comparison. Our results indicate that our
opportunistic routing can exhibit a net improvement in overall
power--delay trade-off over the conventional routing by providing up
to a logarithmic boost in the scaling law. Such a gain is possible
since the receivers can tolerate more interference due to the
increased received signal power provided by the multi-user diversity
gain, which means that having more simultaneous transmissions is
possible.
\end{abstract}

\begin{keywords}
Multi-hop, multi-user diversity, opportunistic routing,
source-destination pair, wireless ad hoc network.
\end{keywords}

\newpage


\section{Introduction}

In~\cite{GuptaKumar:00}, Gupta and Kumar introduced and studied the
throughput scaling in large wireless ad hoc networks. They showed
that a total throughput scaling of $\Theta(\sqrt{n/ \log n})$
[bps/Hz] can be obtained by using a multi-hop strategy when $n$
source-destination (S--D) pairs are randomly distributed in a unit
area.\footnote{We use the following notations: i) $f(x)=O(g(x))$
means that there exist constants $M$ and $m$ such that $f(x)\le
Mg(x)$ for all $x>m$. ii) $f(x)=o(g(x))$ means that
$\underset{x\rightarrow\infty}\lim\frac{f(x)}{g(x)}=0$. iii)
$f(x)=\Omega(g(x))$ if $g(x)=O(f(x))$. iv) $f(x)=\omega(g(x))$ if
$g(x)=o(f(x))$. v) $f(x)=\Theta(g(x))$ if $f(x)=O(g(x))$ and
$g(x)=O(f(x))$~\cite{Knuth:76}.} Multi-hop schemes were then further
developed and analyzed in the
literature~\cite{GuptaKumar:03,DousseFranceschettiThiran:06,FranceschettiDouseTseThiran:07,XueXieKumar:05,NebatCruzBhardwaj:09,ElGamalMammenPrabhakarShah:06,ElGamalMammenPrabhakarShah:06v2,ElGamalMammen:06}
while their throughput per S--D pair scales far less than
$\Theta(1)$. Recent
studies~\cite{OzgurLevequeTse:07,NiesenGuptaShah:07} have shown that
we can actually achieve $\Theta(n^{1-\epsilon})$ scaling for an
arbitrarily small $\epsilon>0$, i.e., an arbitrarily linear scaling
of the total throughput, by using a hierarchical cooperation
strategy, thereby achieving the best result we can hope for.

Besides the studies to improve the throughput up to the linear
scaling, an important factor that we need to consider in practical
wireless networks is the presence of multi-path fading. The effect
of fading on the scaling laws was studied
in~\cite{JovicicViswanathKulkarni:04,XueXieKumar:05,NebatCruzBhardwaj:09,GuptaKumar:03,ToumpisGoldsmith:04},
where it is shown that achievable scaling laws do not change
fundamentally if all nodes are assumed to have their own traffic
demands (i.e., there are $n$ S--D
pairs)~\cite{JovicicViswanathKulkarni:04,XueXieKumar:05,NebatCruzBhardwaj:09}
or the effect of fading is averaged
out~\cite{JovicicViswanathKulkarni:04,XueXieKumar:05,GuptaKumar:03},
while it is found in~\cite{ToumpisGoldsmith:04} that the presence of
fading can reduce the achievable throughput up to $\log n$. However,
fading can be beneficial by utilizing the multi-user diversity (MUD)
gain provided by the randomness of fading in multi-user
environments, e.g., opportunistic scheduling~\cite{KnoppHumblet:95},
opportunistic beamforming~\cite{ViswanathTseLaroia:02}, and random
beamforming~\cite{SharifHassibi:05} in broadcast channels. Scenarios
exploiting the opportunistic gain also studied in cooperative
networks by applying an opportunistic two-hop relaying
protocol~\cite{CuiHaimovichSomekhPoor:07} and in cognitive radio
networks with opportunistic scheduling~\cite{ShenFitz:09}.
In~\cite{GowaikarHochwaldHassibi:06,EbrahimiMaddah-AliKhandani:07},
strategies for improving the throughput scaling over non-faded
environments were shown in wireless network models that do not
incorporate geometric path loss. In~\cite{Chung:06}, it was shown
how fading improves the throughput using opportunistic routing when
a single active S--D pair exists in a wireless ad hoc network.

In this paper, we analyze the benefits of fading by utilizing
opportunistic routing in multi-hop transmissions when there are
multiple randomly located S--D pairs in a large wireless ad hoc
network. Our routing protocol describes how multiple nodes perform
opportunistic routing simultaneously in a massive scale. To our
knowledge, such an attempt for the network model has never been
conducted in the literature. Since the throughput scaling of a
multi-hop protocol is far less than linear, it is natural to assume
that only a subset of S--D pairs are active at a time and active
S--D pairs are chosen in a round robin fashion. In this paper, we
consider a general scenario where the the number of active S--D
pairs scales as a function of $n$. We are interested in improving
the number of simultaneously supportable S--D pairs to maintain a
constant throughput per S--D pair by using opportunistic routing.

In most network applications, power and delay are also key
performance measures along with the throughput. The trade-off among
these measures has been examined in terms of scaling laws in some
papers~\cite{NeelyModiano:05,ElGamalMammenPrabhakarShah:06,ElGamalMammenPrabhakarShah:06v2,ElGamalMammen:06}.
In this paper, we analyze a power--delay--throughput trade-off of
both opportunistic routing and regular multi-hop routing as the
number of S--D pairs increases up to the operating maximum, while
per-node transmission rate is set to a constant. We first show the
existence of a fundamental trade-off between the total transmission
power consumed by all hops per S--D pair, the average number of hops
per S--D pair, i.e., delay, and the number of active S--D pairs,
which is proportional to the total throughput since we assume
per-node transmission rate is a constant. It is investigated whether
power can be reduced at the expense of increased delay for both
routing scenarios, but a net improvement in overall power--delay
trade-off can be obtained with opportunistic routing. The
improvement comes from the MUD gain over the conventional multi-hop
routing. This increases the average received signal power, which in
turn makes it possible to have more simultaneous transmissions since
more interference is tolerated while per-node transmission rate is
maintained. More specifically, we show that such an MUD gain leads
to a logarithmic performance improvement.

The rest of this paper is organized as follows. Section
\ref{SEC:System} describes our system and channel models. In
Section~\ref{SEC:routing}, our protocols with and without
opportunistic routing are described. In Section~\ref{SEC:scaling},
the power--delay--throughput trade-off for these protocols is
analyzed and compared. Finally, Section \ref{SEC:Conc} summarizes
the paper with some concluding remarks.


\section{System and Channel Models} \label{SEC:System}

We consider a two-dimensional wireless network that consists of $n$
nodes uniformly and independently distributed on a square of unit
area (i.e., dense
network~\cite{GuptaKumar:00,ElGamalMammenPrabhakarShah:06,ElGamalMammenPrabhakarShah:06v2,OzgurLevequeTse:07}).
We randomly pick S--D pairings such that each node is the
destination of exactly one source. We assume that there are $M(n)$
randomly located S--D pairs, which can be active simultaneously,
where $M(n)$ scales slower than $n$. Note that $M(n)$ sources can
generate their own data traffic at the same time.

In this paper, to utilize the opportunistic gain, we adopt the
physical channel model that can capture opportunism by modeling a
realistic fading. The received signal $y_k$ at node $k \in
\{1,\cdots,n\}$ at a given time instance is then given by
\begin{equation}
y_k=\sum_{i\in I}h_{ki}x_i+n_k, \label{EQ:signal} \nonumber
\end{equation}
where $x_i\in\mathcal{C}$ is the signal transmitted by node $i$,
$n_k$ is the circularly symmetric complex additive-white Gaussian
noise with zero mean and variance $N_0$, and $I \subset
\{1,\cdots,n\}$ is the set of simultaneously transmitting nodes. The
channel gain $h_{ik}$ is given by
\begin{equation}
h_{ki}=\frac{g_{ki}}{r_{ki}^{\alpha/2}}, \label{EQ:hki}
\end{equation}
where $g_{ki}$ is the complex fading process between nodes $i$ and
$k$, which is assumed to be Rayleigh with $E[|g_{ki}|^2]=1$ and
independent for different $i$'s and $k$'s. Moreover, we assume the
block fading model, where $g_{ki}$ is constant during one packet
transmission and changes to a new independent value for the next
transmission. $r_{ki}$ and $\alpha>2$ denote the distance between
nodes $i$ and $k$ and the path-loss exponent, respectively. We
assume that the channel state information (CSI) is available at all
receivers, but not at the transmitters.

Throughout this paper $E[\cdot]$ denotes the expectation. Unless
otherwise stated, all logarithms are assumed to be to the base 2.


\section{Routing Protocols} \label{SEC:routing}

In this section, we describe our routing protocols with and without
opportunistic routing. We simply use a multi-hop strategy in both
cases using the nodes other than S--D pairs as relays. Hence, we do
not assume the use of any sophisticated multi-user detection schemes
at the receiver.\footnote{If $M(n)$ scales between $\log n$ and
$n^{1/2-\epsilon}$ for an arbitrarily small $\epsilon>0$, which is
the operating regimes in our work, then multi-hop protocols are
enough to satisfy the order optimality in dense networks (the
detailed proof is not shown in this paper).}

Next let us introduce the scaling parameters $P(n)$ and $D(n)$. The
average number of hops per S--D pair is interpreted as the average
delay and is denoted as $D(n)$. The parameter $P(n)$ denotes the
average total transmit power used by all hops for an S--D pair.
Assuming the transmit power is the same for each hop, we see that
$P(n)$ is equal to $D(n)$ times the transmit power per hop. Since
there is no CSI at the transmitter, we assume that each source node
transmits data to its destination at a fixed target rate $R>0$
independent of $n$. A similar assumption was also made in some
earlier
works~\cite{GuptaKumar:00,GuptaKumar:03,DousseFranceschettiThiran:06,FranceschettiDouseTseThiran:07,XueXieKumar:05,NebatCruzBhardwaj:09,ElGamalMammenPrabhakarShah:06,ElGamalMammenPrabhakarShah:06v2,ElGamalMammen:06,OzgurLevequeTse:07,NiesenGuptaShah:07,JovicicViswanathKulkarni:04,ToumpisGoldsmith:04}.
As in the earlier
studies~\cite{KnoppHumblet:95,ViswanathTseLaroia:02,SharifHassibi:05,CuiHaimovichSomekhPoor:07,ShenFitz:09}
dealing with opportunism under the block fading model, we suppose
that a packet is decoded successfully if the received
signal-to-interference-and-noise ratio (SINR) exceeds a
pre-determined threshold $\eta>0$, which is independent of $n$,
i.e., $\log (1+\text{SINR})\geq R=\log (1+\eta)$. Then, the total
throughput $T(n)$ of the network would be given by $\Theta(M(n))$ if
no transmission fails i.e., there is no outage. In addition, we
scale all the transmit power such that the average total
interference power from the set $I\subset\{1,\cdots,n\}$, consisting
of simultaneously transmitting nodes, is given by $\Theta(1)$. Note
that this strategy does not affect the trade-off among the orders of
power $P(n)$, delay $D(n)$, and total throughput $T(n)$ (see
Section~\ref{SEC:scaling_opp} for more detailed description).

\subsection{Opportunistic Routing}

Opportunistic routing was originally introduced
in~\cite{BiswasMorris:03,ZorziRao:03} and further developed in
various network
scenarios~\cite{ShahWietholterWoliszRabaey:05,ZhongNelakuditi:07,ZengLouZhai:08,BhorkarNaghshvarJavidiRao:09}.
When a packet is sent by a transmitter node, it may be possible that
there are multiple receivers successfully decoding the packet. Among
relaying nodes that successfully decode the transmitted packet for
the current hop, the one that is closest to the destination becomes
the transmitter for the next hop. Since the packet can travel
farther at each hop using this opportunistic routing, the average
number of hops can be reduced. Note that the existing protocol
in~\cite{BiswasMorris:03,ZorziRao:03,ShahWietholterWoliszRabaey:05,ZhongNelakuditi:07,ZengLouZhai:08,BhorkarNaghshvarJavidiRao:09}
was designed simply for the case where there exists a single S--D
pair, and thus it did not incorporate interference between links,
which is a critical problem in wireless networks.

We modify this routing to apply it to our network composed of
multiple nodes performing opportunistic routing simultaneously in a
massive scale. Then, we need to carefully design a routing protocol
while solving the interference problem caused by simultaneously
transmitting nodes. The per-hop distance of this opportunistic
transmission is random. However, we can make sure that there are
multiple successfully receiving nodes in a given square cell with
high probability (whp) if we control the size of the cell and the
distance between the transmitter and the cell. Then, one of the
successfully receiving nodes can be the transmitter for the next
hop. Short signaling messages~\cite{BiswasMorris:03,ZorziRao:03}
need to be exchanged between some candidate relaying nodes and the
corresponding transmit node in order to decide who will be the
transmitter for the next hop.\footnote{Alternatively, a timer-based
strategy can be used for selecting the transmitter for the next
hop~\cite{MaSunZhaoLiu:08}.} These messages are transmitted using a
different time slot from that for data packets. More specifically,
it is assumed that the two different messages are transmitted at
even and odd time slots, respectively, which causes only a factor 2
loss in performance, thus resulting in no degradation in terms of
scaling laws.\footnote{Since our aim is to study the performance in
the limit of infinite packet length under the block fading model, if
the packet length scales fast enough in $n$, then we may conclude
that these signaling messages have a negligible overhead.}

\begin{figure}[t!]
  \begin{center}
  \leavevmode \epsfxsize=0.67\textwidth   
  \leavevmode 
  \epsffile{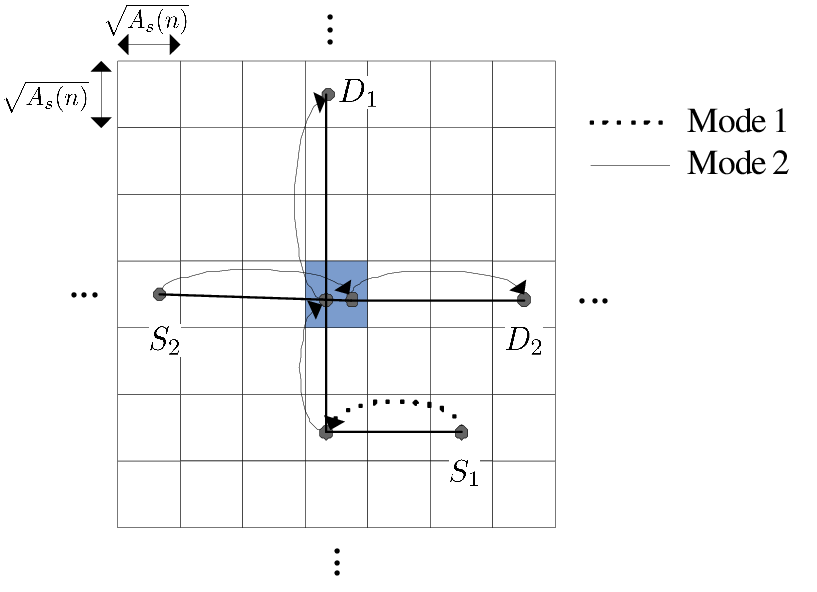}
  \caption{The S-D paths passing through the shaded cell.}
  \label{FIG:SDlines}
  \end{center}
\end{figure}

As shown in Fig.~\ref{FIG:SDlines}, we divide the whole area into
$1/A_s(n)$ square cells with per-cell area $A_s(n)$. Note that
$A_s(n)=\Theta(1/D(n)^2)$ holds since the average distance between
an S--D pair is given by $\Theta(1)$. We assume XY routing, i.e.,
the route for an S--D pair consists of a horizontal and a vertical
paths. Suppose that routing is performed first horizontally and then
vertically for each S--D pair, as illustrated in
Fig.~\ref{FIG:SDlines} ($S_i$ and $D_i$ denote a source and the
corresponding destination node, respectively, for $i=1, 2$). Then,
for each hop in the S--D path, some relay nodes that successfully
decode their packets are selected opportunistically for transmission
in the next hop (the relaying node selection strategy will be
described later in detail). That is, the route for each S--D pair is
not pre-determined. Nodes operate according to the 25-time division
multiple access (TDMA) scheme. This means that the total time is
divided into 25 time slots and nodes in each cell transmit 1/25-th
of the time, while all transmitters in a cell transmit
simultaneously.\footnote{Under our opportunistic routing protocol,
25-TDMA scheme is used 1) to guarantee that there are no transmitter
and receiver nodes near the boundary of two adjacent cells and 2) to
avoid a partitioning problem, which will be discussed later in this
section.} Figure~\ref{FIG:25TDMA} shows an example of simultaneously
transmitting cells depicted as shaded cells.

Our routing protocol consists of two transmission modes, i.e., Modes
1 and 2, where Mode 2 is used for the last two hops to the
destination\footnote{Even for the case where only one hop is needed
between an S--D pair, we can artificially introduce an additional
hop so that there are at least two hops for every S--D pair.} and
Mode 1 is used for all other hops (refer to Fig.~\ref{FIG:SDlines}
for the brief operation of two modes).

\begin{figure}[t!]
  \begin{center}
  \leavevmode \epsfxsize=0.77\textwidth   
  \leavevmode 
  \epsffile{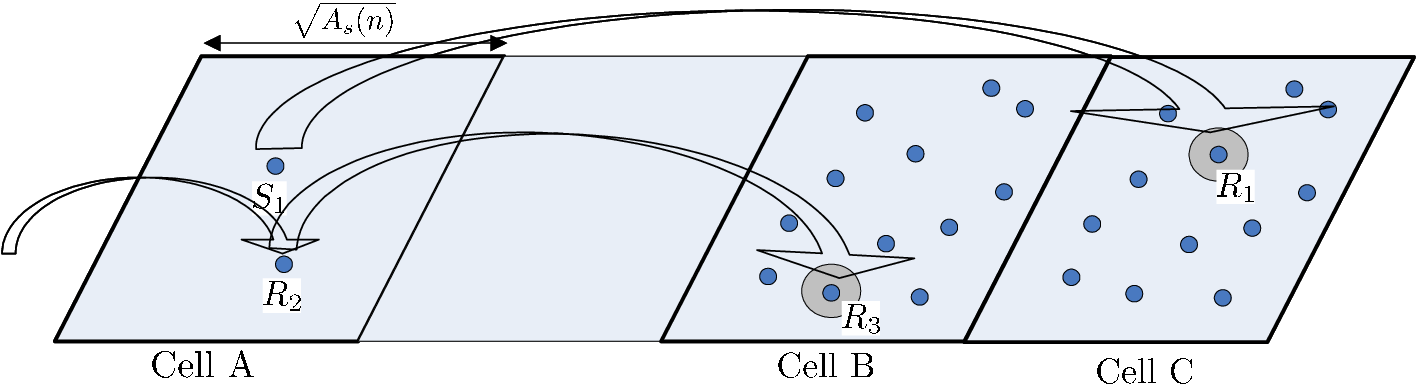}
  \caption{The opportunistic routing protocol in Mode 1.}
  \label{FIG:MUD1}
  \end{center}
\end{figure}

\textbf{Mode 1}: We use an example in Fig.~\ref{FIG:MUD1} to
describe this mode. Transmitting nodes in Cell A transmit packets
simultaneously, where one of those can be either source $S_1$ or
relay node $R_2$. A relay node that successfully decodes the packet
and is two (Cell B) or three (Cell C) cells apart from the
transmitter horizontally (or vertically), for example $R_1$ or $R_3$
in Fig.~\ref{FIG:MUD1}, is arbitrarily chosen for the next hop. If
there is no such node, then an outage occurs, i.e., none of the
nodes satisfy $\text{SINR}\ge \eta$ in the cells. We do not assume
any retransmission scheme in our case since we will make the outage
probability negligibly small. If there are more than one candidate
relay, then we choose one among them arbitrarily. Note that the MUD
gain is roughly equal to the logarithm of the number of nodes in
Cells B and C, which will be rigorously analyzed in the next
section. We perform Mode 1 until the last two hops to the
destination, and then switch to Mode 2. The reason we hop either two
or three cells at a time is because 1) hopping to an immediate
neighbor cell can create huge interference to a receiver node near
the boundary of the two adjacent cells\footnote{By hopping by one
cell, the distance between a receiving node and an interfering node
can be arbitrarily small.} and 2) always hopping by two cells is not
good since it partitions the cells into two groups, even and odd,
and a packet can never be exchanged between the two groups.

\begin{figure}[t!]
  \begin{center}
  \leavevmode \epsfxsize=1.02\textwidth   
  \leavevmode 
  \epsffile{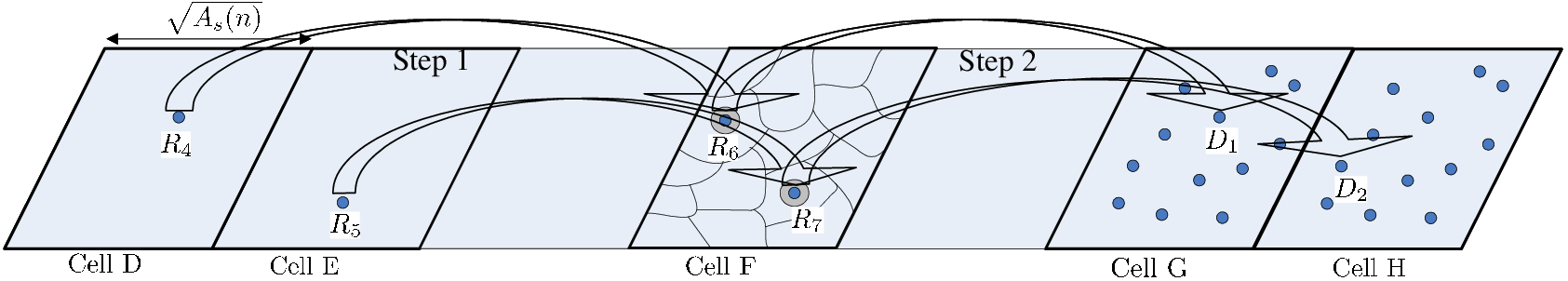}
  \caption{The opportunistic routing protocol in Mode 2.}
  \label{FIG:MUD2}
  \end{center}
\end{figure}

\textbf{Mode 2}: For the last two hops to the destination, Mode 2 is
used. If we use Mode 1 for the last hop, we cannot get any
opportunistic gain since the destination is pre-determined. Hence,
we use the following two-step procedure for Mode 2. We use the
example in Fig.~\ref{FIG:MUD2} to explain this mode.
\begin{itemize}
\item \emph{Step 1:} In this step, a node in Cell D or E (e.g., $R_4$ or $R_5$ in Fig.~\ref{FIG:MUD2})
transmits its packet, whose signal reaches Cell F. This is similar
to what happens in Mode 1 except that we are seeing this from Cell
F's perspective. Assuming $m$ nodes in Cell F, we arbitrarily
partition Cell F into $\sqrt{m}$ sub-cells of equal size, i.e.,
there are roughly $\sqrt{m}$ nodes in each sub-cell. One node is
then opportunistically chosen among the nodes that received the
packet correctly in each sub-cell. Therefore, $\sqrt{m}$ nodes are
chosen in Cell F as potential relays for the packet.
\item \emph{Step 2:}
In Step 2, which corresponds to the last hop, the final destination
in Cell G or H (e.g., $D_1$ or $D_2$ in Fig.~\ref{FIG:MUD2}) sends a
probing packet, i.e., short signaling message, to see which one of
the $\sqrt{m}$ selected relaying nodes in each cell will be the
transmitter for the next hop whose channel link guarantees a
successful packet transmission. Finally, the packet from the
selected relaying node in cell F is transmitted to the final
destination.
\end{itemize}
Although there are only $\sqrt{m}$ candidate nodes in each cell in
Mode 2, whereas there were $m$ nodes in Mode 1, this does not affect
the scaling law since the MUD gain is logarithmic in $m$ and $\log
\sqrt{m}=\frac{1}{2}\log m$.

\subsection{Non-Opportunistic Routing}

In this case, a plain multi-hop
transmission~\cite{GuptaKumar:00,ElGamalMammenPrabhakarShah:06} is
performed with a pre-determined path for each S--D pair consisting
of a source, a destination, and a set of relaying nodes. Therefore,
there is no opportunistic gain. The whole area is also divided into
$1/A_s(n)$ cells with per-cell area $A_s(n)$ and one transmitter in
a cell is arbitrarily chosen while transmitting at a fixed data rate
$R>0$ independent of $n$. We assume the shortest path routing and
the 9-TDMA scheme as
in~\cite{GuptaKumar:00,ElGamalMammenPrabhakarShah:06}. However, even
if interferences are carefully controlled, a transmission may fail
due to fading, causing outages. In this paper, we simply assume that
for the event that an outage occurs (i.e., $\log(1+\text{SINR})<R$)
for a certain hop, such an event is not counted as outage, which
will give an upper bound on the performance.


\section{Power--Delay--Throughput Trade-off}
\label{SEC:scaling}

Our goal in this section is to analyze the power--delay--throughput
trade-offs with and without opportunistic routing. Provided that
per-node transmission rate $R>0$ is given by a constant independent
of $n$, we will show later that there exists a trade-off among
scaling parameters $M(n)$, $P(n)$, and $D(n)$ for the two routing
protocols we consider. By assuming the per-node rate of $R$, the
trade-off among the four parameters $T(n)$, $M(n)$, $P(n)$, and
$D(n)$ is thus essentially reduced to the trade-off among the three
parameters $M(n)$, $P(n)$, and $D(n)$ such that any one of them can
be changed freely, which in turn determines the other two. Note that
with a constant rate $R$, the parameter $M(n)$ is proportional to
the total throughput $T(n)$ since $T(n)=\Theta(M(n))$ if there is no
outage. Note that different protocols will lead to different
power--delay--throughput trade-offs.

If more power is available, then per-hop distance can be extended.
Since the path-loss exponent $\alpha$ is greater than or equal to 2,
the required power increases at least quadratically in the per-hop
distance. On the other hand, the total power consumption of
multi-hop is linear in the number of hops per S--D pair. Therefore,
it seems advantageous to transmit to the nearest neighbor nodes if
we want to minimize the total power. However, this comes at the cost
of increased delay due to more hops. In the following subsections,
we first show that there exists a fundamental trade-off between the
total transmission power consumption per S--D pair, the average
delay per S--D pair, and the total throughput, and then show that
there is a net improvement in the overall power--delay tradeoff when
our opportunistic routing is utilized in the network.

\subsection{Opportunistic Routing}   \label{SEC:scaling_opp}

The relationship among the three parameters $M(n)$, $P(n)$, and
$D(n)$ is derived under the opportunistic routing described above.
More specifically, we are interested in how many S--D pairs, denoted
by $M(n)$, can be active simultaneously while maintaining a constant
transmission rate $R$ per S--D pair. In the following, we mainly
focus on Mode 1 since Mode 2 can be similarly analyzed with a slight
modification. First let $\text{SINR}_{k(m,l_m)}^{l_m}$ denote the
SINR value seen by receiver $k(m,l_m)$ for the $l_m$-th hop of the
$m$-th S--D pair, where $l_m\in\mathcal{H}_m$ and
$m\in\{1,\cdots,M(n)\}$. Here, $\mathcal{H}_m=\{1, 2, \cdots, d_m
D(n)\}$ denotes the set of hops for the $m$-th S--D path, where
$d_m$ is a positive parameter that scales as $O(1)$. Let $N_c(n)$
denote the number of nodes in each cell. Then it follows that
$k(m,l_m)\in\{1,\cdots,2N_c(n)\}$ since two cells are taken into
account for selecting one receiver node. Note that for each hop, a
receiver is either two or three cells apart from the transmitter.
Then, we have
\begin{equation}
\text{SINR}_{k(m,l_m)}^{l_m}=\frac{P_r^{k(m,l_m)}}{N_0+P_I^{k(m,l_m)}},
\label{EQ:SINR1}
\end{equation}
where $P_r^{k(m,l_m)}$ and $P_I^{k(m,l_m)}$ denote the received
signal power at node $k(m,l_m)$ from the desired transmitter
$i(m,l_m)$ for the $l_m$-th hop of the $m$-th S--D pair and the
total interference power at node $k(m,l_m)$ from all interfering
nodes, respectively. Specifically, they are given by
\begin{equation}
P_r^{k(m,l_m)}=|h_{k(m,l_m)i(m,l_m)}|^2\frac{P(n)}{D(n)}
\label{EQ:Prdef}
\end{equation}
and
\begin{equation}
P_I^{k(m,l_m)}=\sum_{i'\in
I\setminus\{i(m,l_m)\}}|h_{k(m,l_m)i'}|^2\frac{P(n)}{D(n)},
\label{EQ:PIdef}
\end{equation}
respectively. Here, $I\subset\{1,\cdots,n\}$ is the set of
simultaneously transmitting nodes. Before establishing our trade-off
results, we start from the following lemma, which shows lower and
upper bounds on the number $N_c(n)$ of nodes in each cell available
as potential relays.

\begin{lemma} \label{LEM:nodenum}
If $A_s(n)=\omega(\log n/n)$, then $N_c(n)$ is between
$((1-\delta_0)A_s(n)n, (1+\delta_0)A_s(n)n)$, i.e.,
$\Theta(A_s(n)n)$, whp for a constant $0<\delta_0<1$ independent of
$n$.
\end{lemma}

The proof of this lemma is given in~\cite{OzgurLevequeTse:07}. In a
similar manner, the number of nodes inside each sub-cell defined in
Mode 2 is between $((1-\delta_0)\sqrt{A_s(n)n},
(1+\delta_0)\sqrt{A_s(n)n})$ whp. We now turn our attention to
quantifying the amount of interference in our schemes in the
following two lemmas.

\begin{lemma} \label{LEM:SDlines}
If $D(n)=o(\sqrt{n/\log n})$ and $D(n)=o({\delta_1}^{M(n)/D(n)})$
for a sufficiently small $\delta_1>1$, then the number of S--D paths
passing through each cell simultaneously is given by
$\Theta(M(n)/D(n))$ whp.
\end{lemma}

\begin{proof}
This proof technique is similar to that
of~\cite{ElGamalMammenPrabhakarShah:06}, but a more general result
is provided for the case where the size of each cell (or
equivalently the average delay $D(n)$) can be controlled
systematically and $M(n)$ scales as a function of $n$. Let
$C_m^{\beta}$ denote an indicator function whose value is one if the
path of the $m$-th S--D pair passes through a fixed cell $\beta$ and
is zero otherwise where $m\in\{1,\cdots,M(n)\}$ and
$\beta\in\{1,\cdots,1/A_s(n)\}$. The total number of paths passing
through the cell $\beta$ is given by
$C^{\beta}=\sum_{m=1}^{M(n)}C_m^{\beta}$, which is the sum of $M(n)$
independent and identically distributed (i.i.d.) Bernoulli random
variables with probability
\begin{equation}
\Pr\left\{C_m^{\beta}=1\right\}=E\left[C_m^{\beta}\right]=\Theta(D(n)A_s(n)),
\nonumber
\end{equation}
where the expectation is taken over the matching of S--D pairs as
well as the node placement. This is because $M(n)$ S--D pairs are
randomly located with uniform distribution on the unit square.
Hence, for any constant $0<\delta_2\le2e-1$, we get the following:
\begin{equation}
P\{C^{\beta}>(1+\delta_2)E[C^{\beta}]\}\le\exp\left(\frac{-E\left[C^{\beta}\right]\delta_2^2}{4}\right)
\nonumber
\end{equation}
from the Chernoff bound~\cite{MotwaniRaghavan:95}. By computing the
following expectation
\begin{equation}
E\left[C^{\beta}\right]=c_0 M(n)D(n)A_s(n)=c_1\frac{M(n)}{D(n)},
\nonumber \label{EQ:ECl}
\end{equation}
where $c_0$ and $c_1$ are some positive constants independent of
$n$, we have
\begin{equation}
P\{C^{\beta}\le(1+\delta_2)E[C^{\beta}]\}\ge
1-\exp\left(-\frac{c_1\delta_2^2}{4}\frac{M(n)}{D(n)}\right).
\nonumber
\end{equation}
Similarly, by the Chernoff bound~\cite{MotwaniRaghavan:95}, it
follows that
\begin{equation}
P\{C^{\beta}\ge(1-\delta_2)E[C^{\beta}]\}\ge
1-\exp\left(-\frac{c_1\delta_2^2}{2}\frac{M(n)}{D(n)}\right),
\nonumber
\end{equation}
thereby yielding
\begin{equation}
P\{(1-\delta_2)E[C^{\beta}]\le
C^{\beta}\le(1+\delta_2)E[C^{\beta}]\}\ge
1-2\exp\left(-\frac{c_1\delta_2^2}{4}\frac{M(n)}{D(n)}\right).
\nonumber
\end{equation}
Due to the fact that there are $1/A_s(n)$ cells in the network, by
applying the union bound over $1/A_s(n)$ cells, it follows that the
number of S--D paths passing through each cell is between
$(c_1(1-\delta_2)M(n)/D(n), c_1(1+\delta_2)M(n)/D(n))$ with
probability of at least
\begin{equation}
1-c_2D(n)^2\exp\left(-\frac{c_1\delta_2^2}{4}\frac{M(n)}{D(n)}\right)
\nonumber
\end{equation}
for constant $c_2>0$ independent of $n$. This tends to one as
$\delta_1^{M(n)/D(n)}/D(n)$ goes to infinity, i.e.,
$D(n)\delta_1^{-M(n)/D(n)}=o(1)$, where $\delta_1$ is a constant
satisfying $1<\delta_1<e^{c_1\delta_2^2/8}$. This completes the
proof of this lemma.
\end{proof}

Using the result of Lemma \ref{LEM:SDlines}, we upper-bound the
total interference as a function of three parameters $M(n)$, $P(n)$,
and $D(n)$ in the following lemma.

\begin{lemma} \label{LEM:interference}
Suppose $D(n)=o(\sqrt{n/\log n})$,
$D(n)=o(n^{-1}\delta_3^{M(n)/D(n)})$, and $\alpha>2$, where
$\delta_3>1$ is a sufficiently small constant. When the 25-TDMA
scheme is used, the total interference power $P_I^{k}$ at receiver
node $k$ from simultaneously transmitting nodes is given by
\begin{equation}
O(P(n)M(n)D(n)^{\alpha-2}) \nonumber
\end{equation}
with probability of at least
\begin{equation}
1-nD(n)\delta_3^{\frac{M(n)}{D(n)}}\exp\left(-c_3\frac{M(n)}{D(n)}\right)
\label{EQ:AppPI}
\end{equation}
for constant $c_3>0$ independent of $n$. Equation (\ref{EQ:AppPI})
tends to one as $n$ increases and the expectation $E[P_I^k]$ of
$P_I^k$ is given by
\begin{equation}
E\left[P_I^k\right]=\Theta\left(P(n)M(n)D(n)^{\alpha-2}\right).
\label{EQ:EPI} \nonumber
\end{equation}
\end{lemma}

The proof of this lemma is presented in
Appendix~\ref{PF:interference}. Note that $P_I^{k}$ depends on the
path loss exponent $\alpha$.

Now to simply find a lower bound on the throughput, suppose that the
threshold value $\eta$ is set to 1.\footnote{If $\eta$ is optimized,
then the achievable rates can be slightly improved. However, for
analytical convenience, we just assume $\eta=1$.} Let us focus on
the $m$-th S--D pair, where $m\in\{1,\cdots,M(n)\}$. Note that a
packet from the $m$-th source passes through the $m$-th S--D pair's
routing path that consists of the set
$\mathcal{H}_m=\{1,2,\cdots,d_m D(n)\}$ of hops. Accordingly, if the
condition $\text{SINR}\ge 1(=\eta)$ is not guaranteed for at least
one among $d_m D(n)$ hops of the path, then the data transmission
for the $m$-th S--D pair will fail, causing outages. To analyze the
achievable throughput, it is thus important to examine the
probability that the source's packet is successfully delivered to
the final destination node while satisfying $\text{SINR}\ge 1$ for
all hops $l_m\in\mathcal{H}_m$. To be concrete, let $A_m$ denote the
event that there is no outage for the $m$-th S--D pair, the
condition $\text{SINR}_{k(m,l_m)}^{l_m}\ge 1$ holds for at least one
receiver $k(m,l_m)\in\{1,\cdots,2N_c(n)\}$ per hop for all hops
$l_m\in\mathcal{H}_m$ of the $m$-th S--D pair. Since the Gaussian is
the worst additive noise~\cite{Medard:00,DiggaviCover:01}, assuming
it lower-bounds the capacity. Hence, by assuming full CSI at the
receiver side, the total throughput $T(n)$ is then given by
\begin{eqnarray} T(n)\!\!\!\!\!\!\!&&\ge
 \sum_{m=1}^{M(n)}\Pr\left\{A_m\right\} \nonumber\\
&&\ge
\sum_{m=1}^{M(n)}\Biggl\{1-\sum_{l_m=1}^{d_mD(n)-2}\left(\Pr\left\{\text{SINR}_{k(m,l_m)}^{l_m}<1\right\}\right)^{2(1-\delta_0)A_s(n)n}
\nonumber\\ && \quad
-(1+\delta_0)\sqrt{A_s(n)n}\left(\Pr\left\{\text{SINR}_{k(m,l_m)}^{d_m
D(n)-1}<1\right\}\right)^{(1-\delta_0)\sqrt{A_s(n)n}} \nonumber\\&&
\quad-\left(\Pr\left\{\text{SINR}_{k(m,l_m)}^{d_m
D(n)}<1\right\}\right)^{(1-\delta_0)\sqrt{A_s(n)n}}\Biggr\}
\nonumber\\
&&\ge
\sum_{m=1}^{M(n)}\Biggl\{1-\sum_{l_m=1}^{d_mD(n)-2}\left(\Pr\left\{\text{SINR}_{k(m,l_m)}^{l_m}<1\right\}\right)^{(1-\delta_0)\sqrt{A_s(n)n}}
\nonumber\\ && \quad
-(1+\delta_0)\sqrt{A_s(n)n}\left(\Pr\left\{\text{SINR}_{k(m,l_m)}^{d_m
D(n)-1}<1\right\}\right)^{(1-\delta_0)\sqrt{A_s(n)n}} \nonumber\\&&
\quad-\left(\Pr\left\{\text{SINR}_{k(m,l_m)}^{d_m
D(n)}<1\right\}\right)^{(1-\delta_0)\sqrt{A_s(n)n}}\Biggr\}.
\label{EQ:Tnlower}
\end{eqnarray}
Here, the first inequality comes from the fact that per-node
transmission rate is given by $R=\log(1+\eta)=1$. The second
inequality holds by applying the union bound over all hops for each
S--D pair, where the set $\mathcal{H}_m$ of hops is specified by
$l_m\in\{1,\cdots,d_m D(n)-2\}$ for Mode 1 and the last two hops to
the destination (i.e., $l_m\in\{d_m D(n)-1, d_m D(n)\}$) for Mode 2.
In order to further compute the right-hand side of
(\ref{EQ:Tnlower}), we need to know the distribution of the SINR,
which is difficult to obtain for a general class of channel models
consisting of both geometric and fading effects. Instead,
in~\cite{WeberAndrewsJindal:07}, asymptotic upper and lower bounds
on the cumulative distribution function (cdf) of SINR were
characterized.

In this paper, as mentioned earlier, we assume that the average
total interference power $E\left[P_I^{k(m,l_m)}\right]$ at receiver
node $k(m,l_m)$ ($l_m\in\mathcal{H}_m$ and $m\in\{1,\cdots,M(n)\}$)
is $\Theta(1)$, which is the best situation we can hope for to
maintain the fixed transmission rate $R>0$ for each hop. This is
because if $E\left[P_I^{k(m,l_m)}\right]$ is not $O(1)$, then we can
scale down all transmit powers proportionally such that
$E\left[P_I^{k(m,l_m)}\right]=\Theta(1)$ without loss of optimality
in scaling. This is because the received signal power
$P_r^{k(m,l_m)}$ from the desired transmitter should be
$\Theta\left(E\left[P_I^{k(m,l_m)}\right]\right)$ to maintain a
fixed rate $R$ per S--D pair and having such higher power from both
the signal and the interference is unnecessary. On the other hand,
if $E\left[P_I^{k(m,l_m)}\right]=o(1)$, then it follows that
$\text{SINR}_{k(m,l_m)}^{l_m}=\Theta\left(P_r^{k(m,l_m)}\right)$. We
can thus scale up all transmit powers proportionally such that
$E\left[P_I^{k(m,l_m)}\right]=\Theta(1)$ in order to increase the
SINR value, resulting in improved per-node transmission rate. Then
using Lemma~\ref{LEM:interference} and the fact that
$E\left[P_I^{k(m,l_m)}\right]=\Theta(1)$, we obtain
\begin{equation}
P(n)M(n)D(n)^{\alpha-2}=\Theta(1) \label{EQ:PI1}
\end{equation}
for receiver node $k(m,l_m)$. This assumption makes the analysis of
scaling laws much simpler.
As a consequence, it is possible to find the cdf of the SINR in
(\ref{EQ:Tnlower}) when our opportunistic routing is utilized. Let
$B$ denote the event that $P_I^{k}=O(1)$ holds for receiver node $k$
in the network. By using (\ref{EQ:hki})--(\ref{EQ:Prdef}), and the
condition $D(n)=o(n^{-1}\delta_3^{M(n)/D(n)})$ in
Lemma~\ref{LEM:interference}, we then have
\begin{eqnarray}
\Pr\left\{\text{SINR}_{k(m,l_m)}^{l_m}<1\right\}\!\!\!\!\!\!\!&&=\Pr\left\{\frac{|h_{k(m,l_m)i(m,l_m)}|^2P(n)/D(n)}{N_0+P_I^{k(m,l_m)}}<1\right\}
\nonumber\\
&&\le 1-\Pr\left\{\frac{|h_{k(m,l_m)i(m,l_m)}|^2P(n)/D(n)}{N_0+P_I^{k(m,l_m)}}\ge1\biggr|B\right\}\Pr\left\{B\right\} \nonumber\\
&&\le
1-\Pr\left\{\left|h_{k(m,l_m)i(m,l_m)}\right|^2\frac{P(n)}{D(n)}\ge
c_4\right\}\Pr\left\{B\right\}
\nonumber\\
&& =1-\Pr\left\{\left|g_{k(m,l_m)i(m,l_m)}\right|^2\ge
\frac{c_5}{P(n)D(n)^{\alpha-1}}\right\}\Pr\left\{B\right\}
\nonumber\\&&=1-\exp\left(-\frac{c_5}{P(n)D(n)^{\alpha-1}}\right)\Pr\left\{B\right\}
\nonumber\\
&&\le
1-\exp\left(-\frac{c_5}{P(n)D(n)^{\alpha-1}}\right)\left(1-nD(n)\delta_3^{\frac{M(n)}{D(n)}}\exp\left(-c_3\frac{M(n)}{D(n)}\right)\right)
\nonumber\\&&=1-c_6\exp\left(-\frac{c_5}{P(n)D(n)^{\alpha-1}}\right),
\label{EQ:SINRcdf}
\end{eqnarray}
where $c_3$, $c_4$, $c_5$, and $c_6$ denote some positive constants
independent of $n$, and $\delta_3>1$ is a sufficiently small
constant. Here, the second equality comes from the fact that per-hop
distance is given by $\Theta(1/D(n))$. The third equality holds
since the squared channel gain $\left|g_{k(m,l_m)i(m,l_m)}\right|^2$
follows the chi-square distribution with two degrees of freedom. The
third inequality comes from (\ref{EQ:AppPI}) and (\ref{EQ:PI1}). The
last equality holds since it follows that
\begin{equation}
1-nD(n)\delta_3^{\frac{M(n)}{D(n)}}\exp\left(-c_3\frac{M(n)}{D(n)}\right)=\Theta(1)
\nonumber
\end{equation}
under the condition $D(n)=o(n^{-1}\delta_3^{M(n)/D(n)})$. Note that
the upper bound on the probability
$\Pr\left\{\text{SINR}_{k(m,l_m)}^{l_m}<1\right\}$ is identical for
all hops $l_m\in\mathcal{H}_m$ since it does not depend on $l_m$.
Now we are ready to derive the scaling laws for $P(n)$, $D(n)$, and
$T(n)$ in terms of $M(n)$ by using (\ref{EQ:Tnlower}),
(\ref{EQ:PI1}), and (\ref{EQ:SINRcdf}).

\begin{theorem} \label{THM:PDM}
Suppose that $E\left[P_I^k\right]=\Theta(1)$ for receiver node $k$
(i.e., $P(n)M(n)D(n)^{\alpha-2}=\Theta(1)$), a constant rate $R$ per
S--D pair, $D(n)=o(n^{-1}\delta^{M(n)/D(n)})$, and $\alpha>2$, where
$\delta>1$ is a sufficiently small constant. If
$M(n)=O(n^{1/2-\epsilon})$ and $M(n)=\Omega(\log n)$, the
opportunistic routing achieves
\begin{equation}
P(n)=\Theta\left(\frac{M(n)^{-\alpha+1}}{(\log
n)^{-\alpha+2}}\right), \label{EQ:scaling1}
\end{equation}
\begin{equation}
D(n)=\Theta\left(\frac{M(n)}{\log n}\right), \label{EQ:scaling2}
\end{equation}
and the total throughput $T(n)=\Omega(M(n))$ whp, where $\epsilon>0$
is an arbitrarily small constant.
\end{theorem}

\begin{proof}
By substituting (\ref{EQ:SINRcdf}) into (\ref{EQ:Tnlower}), the
total throughput $T(n)$ can be lower-bounded by
\begin{eqnarray}
T(n)\!\!\!\!\!\!\!&&\ge \sum_{m=1}^{M(n)}\Biggl\{1-\left(d_mD(n)+(1+\delta_0)\sqrt{A_s(n)n}\right) \nonumber\\
&&
\quad\cdot\left(1-c_6\exp\left(-\frac{c_5}{P(n)D(n)^{\alpha-1}}\right)\right)^{(1-\delta_0)\sqrt{A_s(n)n}}\Biggr\}
\nonumber\\ &&\ge M(n)\Biggl\{1-\left(\bar{D}(n)+2\sqrt{A_s(n)n}\right) \nonumber\\
&&
\quad\cdot\left(1-c_6\exp\left(-\frac{c_5}{P(n)D(n)^{\alpha-1}}\right)\right)^{(1-\delta_0)\sqrt{A_s(n)n}}\Biggr\},
\nonumber
\end{eqnarray}
where $\bar{D}(n)=\max\{d_1,\cdots,d_{M(n)}\}D(n)$. To guarantee
$T(n)=\Omega(M(n))$ whp with no outage for transmissions, we thus
need the following equality:
\begin{equation}
\left(\bar{D}(n)+2\sqrt{A_s(n)n}\right)\left(1-c_6\exp\left(-\frac{c_5}{P(n)D(n)^{\alpha-1}}\right)\right)^{(1-\delta_0)\sqrt{A_s(n)n}}=\epsilon_0
\label{EQ:desiredsig} \nonumber
\end{equation}
for an arbitrarily small $\epsilon_0>0$. Then, it follows that
\begin{eqnarray}
\max\left\{D(n),\sqrt{\frac{n}{D(n)^2}}\right\}\left(1-c_6\exp\left(-\frac{c_5}{P(n)D(n)^{\alpha-1}}\right)\right)\!\!\!\!\!\!\!\!\!\!&&^{(1-\delta_0)\sqrt{\frac{n}{D(n)^2}}}=\Theta(1),
\nonumber
\end{eqnarray}
which yields
\begin{eqnarray}
\left\{\begin{array}{lll}
P(n)D(n)^{\alpha-1}\log\left(\frac{\sqrt{n}}{D(n)\log
n}\right)=\Theta(1) &\textrm{if
$D(n)=o\left(n^{1/4}\right)$} \\
P(n)D(n)^{\alpha-1}\log\left(\frac{\sqrt{n}}{D(n)\log
D(n)}\right)=\Theta(1) &\textrm{if
$D(n)=\Omega\left(n^{1/4}\right)$.}
\end{array}\right. \label{EQ:desiredPr}
\end{eqnarray}
After some calculation, using (\ref{EQ:PI1}) and
(\ref{EQ:desiredPr}), we obtain
\begin{equation}
M(n)=\Theta\left(\left(P(n)M(n)\right)^{\frac{-1}{\alpha-2}}\log\left(\frac{\sqrt{n}(P(n)M(n))^{\frac{1}{\alpha-2}}}{\log
n}\right)\right) \label{EQ:scaling3} \nonumber
\end{equation}
and
\begin{equation}
M(n)=\Theta\left(D(n)\log\left(\frac{\sqrt{n}}{D(n)\log
n}\right)\right). \label{EQ:scaling4}
\end{equation}
From (\ref{EQ:scaling4}) and the condition
$D(n)=o(n^{-1}\delta^{M(n)/D(n)})$, it follows that
\begin{equation}
D(n)=O\left(\frac{n^{1/2-\epsilon}}{\log n}\right) \label{EQ:Dncon}
\end{equation}
and
\begin{equation}
M(n)=O\left(n^{1/2-\epsilon}\right) \label{EQ:Mncon}
\end{equation}
for an arbitrarily small $\epsilon>0$, and hence, we have
\begin{equation}
M(n)=\Theta\left(\left(P(n)M(n)\right)^{\frac{-1}{\alpha-2}}\log
n\right) \nonumber
\end{equation}
and
\begin{equation}
M(n)=\Theta\left(D(n)\log n\right) \nonumber
\end{equation}
under the constraints (\ref{EQ:Dncon}) and (\ref{EQ:Mncon}), finally
resulting in (\ref{EQ:scaling1}) and (\ref{EQ:scaling2}). Let
$\delta=\min\{\delta_1, \delta_3\}$, where $\delta_1$ and $\delta_3$
are shown in Lemmas \ref{LEM:SDlines} and \ref{LEM:interference},
respectively. If we choose a constant $\delta_4\ge3/(2\log \delta)$
independent of $n$ satisfying
\begin{equation}
\frac{M(n)}{D(n)}=\delta_4\log n, \nonumber
\end{equation}
then it is seen that the condition
$D(n)=o(n^{-1}\delta^{M(n)/D(n)})$ always holds from
(\ref{EQ:scaling2}). We also have $M(n)=\Omega(\log n)$ due to
$D(n)=\Omega(1)$. This completes the proof of this theorem.
\end{proof}

Note that the logarithmic terms in (\ref{EQ:scaling1}) and
(\ref{EQ:scaling2}) are due to the MUD gain of the opportunistic
routing. The operating regimes correspond to the case where the
number $M(n)$ of simultaneously active S--D pairs scales between
$\log n$ and $n^{1/2-\epsilon}$. Furthermore, we see that $P(n)$
monotonically decreases with respect to $M(n)$ while $D(n)$ scales
almost linearly. We finally remark that using (\ref{EQ:scaling1})
and (\ref{EQ:scaling2}) yields the relationship
\begin{equation}
P(n)=\Theta\left(\frac{D(n)^{-\alpha+1}}{\log n}\right)
\label{EQ:tradeoffopp}
\end{equation}
between the two scaling parameters $P(n)$ and $D(n)$.

\subsection{Non-Opportunistic Routing}

In this subsection, the scaling result of non-opportunistic routing
is shown for comparison. As addressed before, the total interference
power $P_I^{k}$ at receiver node $k$ needs to be $O(1)$, and it thus
follows that $P(n)M(n)D(n)^{\alpha-2}=\Theta(1)$ due to
Lemma~\ref{LEM:interference} and (\ref{EQ:PI1}). In this case, we
investigate how the delay $D(n)$ and the power $P(n)$ scale when
there are $M(n)$ simultaneously active S--D pairs, while maintaining
a constant $R>0$, as in Section~\ref{SEC:scaling_opp}. The
power--delay--throughput trade-off is derived in the following
theorem.

\begin{theorem}
Suppose that $P_I^{k}=\Theta(1)$ for each receiver node $k$ (i.e.,
$P(n)M(n)D(n)^{\alpha-2}=\Theta(1)$), a constant rate $R$ per S--D
pair, and $\alpha>2$. If $M(n)=O(n^{1/2-\epsilon})$ and
$M(n)=\Omega(\log n)$ for an arbitrarily small $\epsilon>0$, then
the non-opportunistic routing achieves
\begin{equation}
P(n)=\Theta(M(n)^{-\alpha+1}), \label{EQ:Pnno}
\end{equation}
\begin{equation}
D(n)=\Theta(M(n)), \label{EQ:Dnno}
\end{equation}
and the total throughput $T(n)=\Theta(M(n))$.
\end{theorem}

The proof of this lemma almost follows the same line as that of
Theorem \ref{THM:PDM}. Note that there is no logarithmic term in the
two equations shown above. We also remark that using (\ref{EQ:Pnno})
and (\ref{EQ:Dnno}) results in the relationship
\begin{equation}
P(n)=\Theta(D(n)^{-\alpha+1}) \label{EQ:PDno}
\end{equation}
between $P(n)$ and $D(n)$.

\subsection{Performance Comparison}

\begin{figure}[t!]
  \begin{center}
  \leavevmode \epsfxsize=0.64\textwidth   
  \leavevmode 
  \epsffile{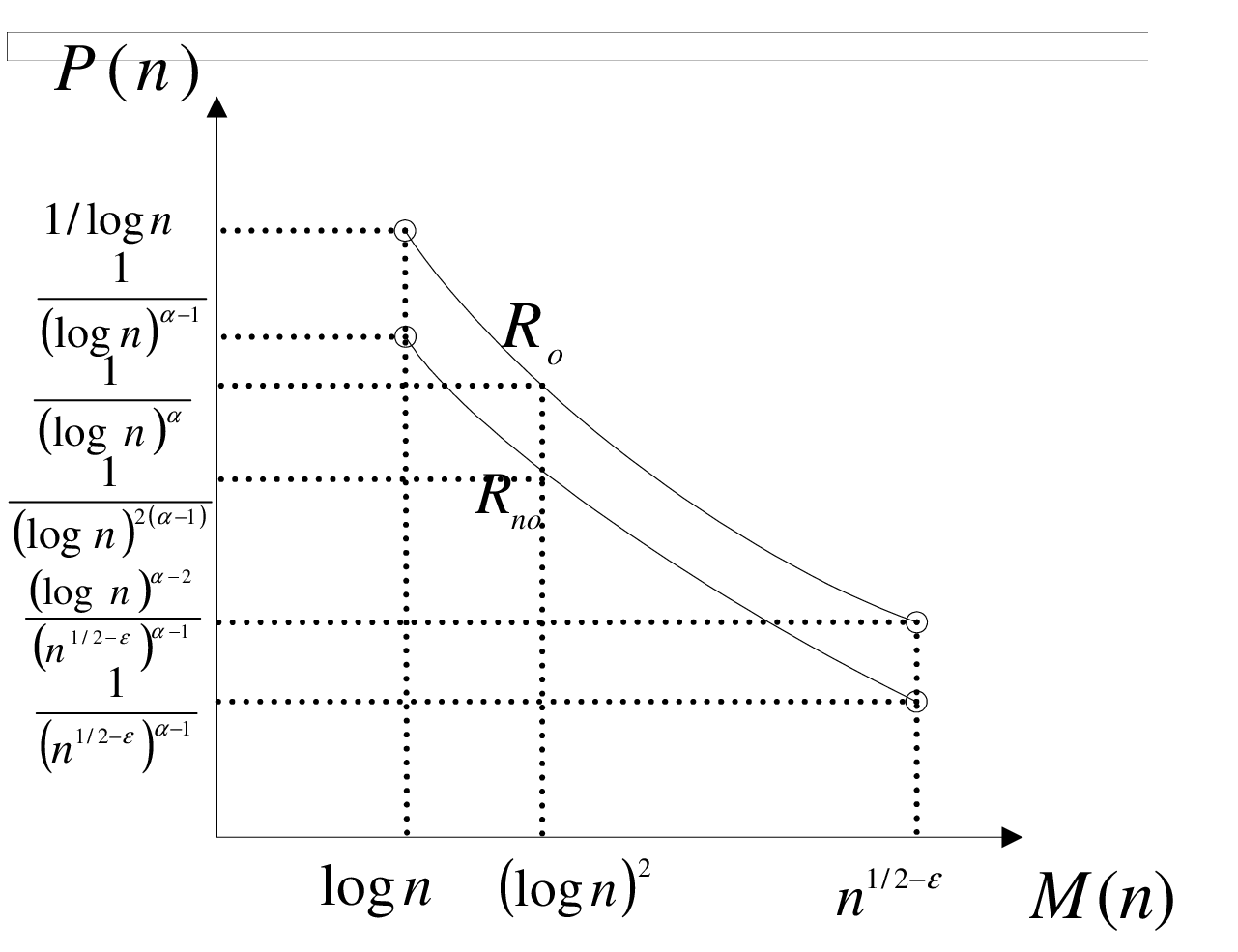}
  \caption{The power scaling with respect to $M(n)$.}
  \label{FIG:power}
  \end{center}
\end{figure}

\begin{figure}[t!]
  \begin{center}
  \leavevmode \epsfxsize=0.64\textwidth   
  \leavevmode 
  \epsffile{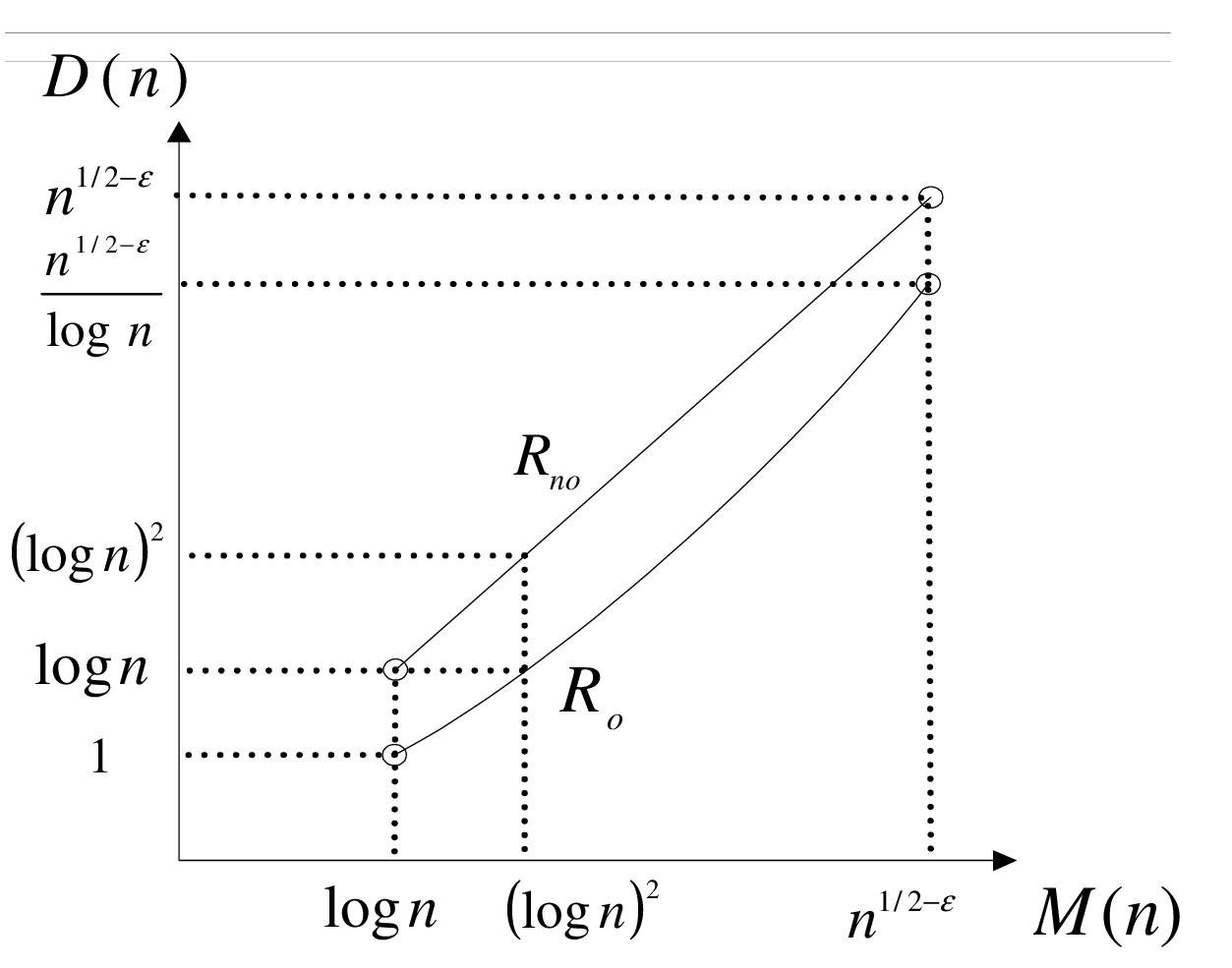}
  \caption{The delay scaling with respect to $M(n)$.}
  \label{FIG:delay}
  \end{center}
\end{figure}

\begin{figure}[t!]
  \begin{center}
  \leavevmode \epsfxsize=0.64\textwidth   
  \leavevmode 
  \epsffile{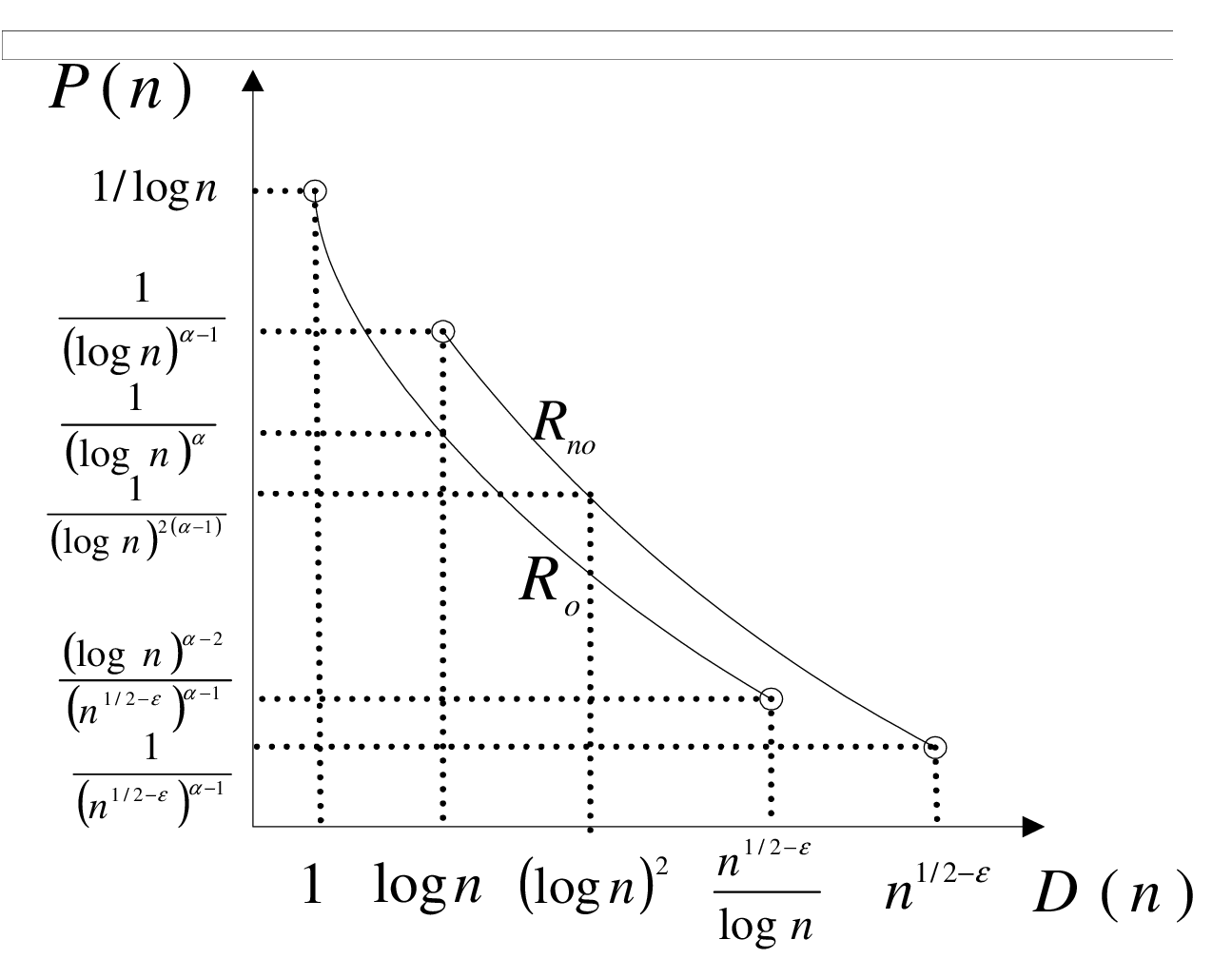}
  \caption{The power--delay trade-off.}
  \label{FIG:tradeoff}
  \end{center}
\end{figure}

Now we show that the opportunistic routing exhibits a net
improvement in overall power--delay trade-off over the conventional
non-opportunistic routing. Figures \ref{FIG:power} and
\ref{FIG:delay} show how the power $P(n)$ and the delay $D(n)$ scale
with respect to the number $M(n)$ of simultaneously active S--D
pairs, corresponding to the total throughput $T(n)$. $R_o$ and
$R_{no}$ denote the scaling curves with and without opportunistic
routing, respectively. We only take into account the range of $M(n)$
between $\log n$ and $n^{1/2-\epsilon}$ for an arbitrarily small
$\epsilon>0$, which is the operating regimes in our work, due to
various constraints we assume in the model. Hence, the MUD gain may
not be guaranteed if $M(n)$ scales faster than $n^{1/2-\epsilon}$
for a vanishingly small $\epsilon>0$ (e.g., it is shown
in~\cite{NebatCruzBhardwaj:09} that when $M(n)=\Theta(\sqrt{n})$,
the benefit of fading cannot be exploited in terms of scaling laws).
We observe that $P(n)$ decreases while $D(n)$ increases as we have
more active S--D pairs in both schemes. This is because we assume a
fixed transmission rate $R>0$ independent of $n$, which implies that
for receiver $k(m,l_m)$ ($l_m\in\mathcal{H}_m$ and
$m\in\{1,\cdots,M(n)\}$), $P_r^{k(m,l_m)}$ and $P_I^{k(m,l_m)}$ need
to be $\Omega(1)$ and $O(1)$, respectively, as mentioned earlier.
Then, to maintain the interference level $P_I^{k(m,l_m)}$ at $O(1)$
as $M(n)$ increases, more hops per S--D pair are needed, i.e.,
per-hop distance is reduced. Hence, from the above argument, we may
conclude that the power is reduced at the expense of the increased
delay, and therefore, there is a fundamental trade-off between the
two scaling parameters $P(n)$ and $D(n)$. Furthermore, it is seen
that utilizing the opportunistic routing increases the power
compared to the non-opportunistic routing case, but it can reduce
the delay significantly. Thus, it is not clear whether our
opportunistic routing is beneficial or not from Figs.
\ref{FIG:power} and \ref{FIG:delay}. However, if we plot the power
$P(n)$ versus the delay $D(n)$ as in Fig.~\ref{FIG:tradeoff}, then
it can be clearly seen that opportunistic routing ($R_o$) exhibits a
better overall power--delay trade-off than that of non-opportunistic
routing scheme ($R_{no}$), while providing a logarithmic boost in
the scaling law. For example, when opportunistic routing is used, if
the delay $D(n)$ is given by $\log n$, then the power $P(n)$ is
reduced by $\log n$. In this case, it is further seen from
Fig.~\ref{FIG:delay} that the number $M(n)$ of simultaneously
supportable S--D pairs is improved by $\log n$, i.e., logarithmic
boost on the total throughput $T(n)$. This gain comes from the fact
that the received signal power increases due to the MUD gain based
on the use of opportunistic routing, which allows more simultaneous
transmissions since more interference can be tolerated.


\section{Conclusion} \label{SEC:Conc}

The scaling behavior of ad hoc networks using opportunistic routing
protocol in the presence of fading has been characterized.
Specifically, it was shown how the power, delay, and total
throughput scale as the number of S--D pairs increases, while
maintaining a constant per-node transmission rate. We proved that
for the range of simultaneously active S--D pairs between $\log n$
and $n^{1/2-\epsilon}$ for an arbitrarily small $\epsilon>0$, the
opportunistic routing exhibits a net improvement in overall
power--delay trade-off over the conventional scheme employing
non-opportunistic routing, while providing up to a $\log n$ boost in
the scaling law due to the MUD gain.


\appendix

\section{appendix}

\subsection{Proof of Lemma \ref{LEM:interference}}   \label{PF:interference}

\begin{figure}[t!]
  \begin{center}
  \leavevmode \epsfxsize=0.60\textwidth   
  \leavevmode 
  \epsffile{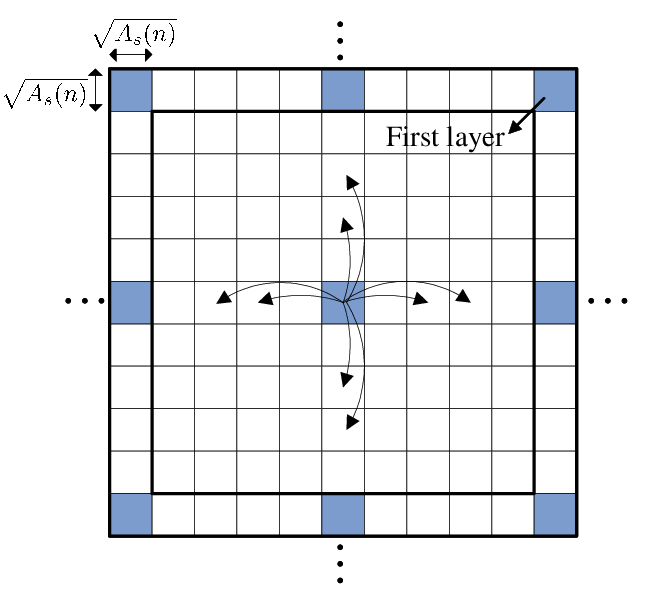}
  \caption{Grouping of interfering cells in the 25-TDMA scheme. The first layer represents the outer eight shaded cells.}
  \label{FIG:25TDMA}
  \end{center}
\end{figure}

There are $8l$ interfering cells in the $l$-th layer of 25-TDMA
(refer to Fig. \ref{FIG:25TDMA}). Let $P_{I,(l)}^{k}$ denote the
total interference power at a fixed receiver node $k$ from
simultaneously transmitting nodes in the $l$-th layer, where
$l\in\{1,\cdots,\bar{l} D(n)\}$ for some constant $\bar{l}>0$
independent of $n$. Note that the distance between a receiving node
and an interfering node in the $l$-th layer is between
$((5l-4)\sqrt{A_s(n)}, 8(5l-4)\sqrt{A_s(n)})$. Suppose that the
Euclidean distance among the links above is given by
$8(5l-4)\sqrt{A_s(n)}$, thereby providing a lower bound for
$E\left[P_{I,(l)}^{k}\right]$. By Lemma \ref{LEM:SDlines}, the
number of simultaneous transmitters in each cell is given by
$\Theta(M(n)/D(n))$ whp. Thus, from (\ref{EQ:hki}) and
(\ref{EQ:PIdef}), the expectation $E[P_I^{(l)}]$ is lower-bounded by
\begin{eqnarray}
E\left[P_{I,(l)}^{k}\right]\!\!\!\!\!\!\!&&\ge \frac{c_7(8l)
P(n)/D(n)}{\left(8(5l-4)\sqrt{A_s(n)}\right)^\alpha}\frac{M(n)}{D(n)}E\left[|g_{ki}|^2\right]
\nonumber\\ &&\ge \frac{c_7}{8^{\alpha-1}(5l-4)^{\alpha-1}}
\frac{P(n)/D(n)}{(\sqrt{A_s(n)})^\alpha}\frac{M(n)}{D(n)}
\nonumber\\&&= \frac{c_8P(n)M(n)D(n)^{\alpha-2}}{(5l-4)^{\alpha-1}}
\nonumber
\end{eqnarray}
for any nodes $i$ and $k$, where $c_7$ and $c_8$ are some positive
constants independent of $n$. Similarly by taking
$(5l-4)\sqrt{A_s(n)}$ for the Euclidean distance between a receiver
and simultaneously transmitting nodes in the $l$-th layer, we obtain
\begin{eqnarray}
E\left[P_{I,(l)}^{k}\right]\!\!\!\!\!\!\!&&\le
\frac{c_9P(n)M(n)D(n)^{\alpha-2}}{(5l-4)^{\alpha-1}}
\label{EQ:PIlupper} \nonumber
\end{eqnarray}
for constant $c_9>0$ independent of $n$, which results in
\begin{equation}
E\left[P_{I,(l)}^{k}\right]=\Theta\left(\frac{P(n)M(n)D(n)^{\alpha-2}}{(5l-4)^{\alpha-1}}\right).
\nonumber
\end{equation}
Moreover, let $P_{I,(0)}^{k}$ be the total interference power from
other transmitting nodes in the cell including a desired transmitter
(see the shaded cell located in the center in Fig.
\ref{FIG:25TDMA}). Then as above, we have
$E\left[P_{I,(0)}^{k}\right]=\Theta(P(n)M(n)D(n)^{\alpha-2})$, and
hence, it follows that
$E\left[P_I^{k}\right]=\Theta(P(n)M(n)D(n)^{\alpha-2})$ since
$\sum_{l}1/(5l-4)^{\alpha-1}$ is bounded by a certain constant for
$\alpha>2$.

Now we focus on computing
$\Pr\left\{P_{I,(l)}^{k}>(1+\delta_5)E\left[P_{I,(l)}^{k}\right]\right\}$
by using the Chernoff bound, where $\delta_5>0$ is a constant
independent of $n$. From the fact that $r_{ki}=\Theta(r_{ki'})$ for
all transmitting nodes $i$ and $i'$ in the same layer and receiving
node $k$, we have
\begin{eqnarray}
&&\Pr\left\{P_{I,(l)}^{k}>(1+\delta_5)E\left[P_{I,(l)}^{k}\right]\right\}
\nonumber\\ && =\Pr\left\{\sum_{i\in
I_l}\frac{|g_{ki}|^2}{r_{ki}^{\alpha}}>(1+\delta_5)E\left[\sum_{i\in
I_l}\frac{|g_{ki}|^2}{r_{ki}^{\alpha}}\right]\right\} \nonumber\\&&
\le\Pr\left\{\sum_{i\in
I_l}\frac{|g_{ki}|^2}{\left(5l-4)\sqrt{A_s(n)}\right)^{\alpha}}>(1+\delta_5)E\left[\sum_{i\in
I_l}\frac{|g_{ki}|^2}{r_{ki}^{\alpha}}\right]\right\} \nonumber \\
&&=\Pr\left\{\sum_{i\in
I_l}|g_{ki}|^2>c_{10}(8l)(1+\delta_5)\frac{M(n)}{D(n)}\right\}
\label{EQ:PIleq}
\end{eqnarray}
for constant $c_{10}>0$ independent of $n$. Here, $I_l$ is the set
of simultaneously interfering nodes in the $l$-th layer. Since the
Chernoff bound for the sum of i.i.d. chi-square random variables
$|g_{ki}|^2$ with two degrees of freedom is given
by~\cite{ShwartzWeiss:95}, for a certain constant
$0<\epsilon_1<\delta_5-\log(1+\delta_5)$, (\ref{EQ:PIleq}) can be
upper-bounded by
\begin{eqnarray}
\Pr\left\{P_{I,(l)}^{k}>(1+\delta_5)E\left[P_{I,(l)}^{k}\right]\right\}\!\!\!\!\!\!\!&&\le
(1+\delta_5)^{c_{10}(8l)\frac{M(n)}{D(n)}}\exp\left(-c_{10}(8l)\frac{M(n)}{D(n)}(\delta_5-\epsilon_1)\right)
\nonumber\\ &&\le
(1+\delta_5)^{8c_{10}\frac{M(n)}{D(n)}}\exp\left(-8c_{10}\frac{M(n)}{D(n)}(\delta_5-\epsilon_1)\right),
\nonumber
\end{eqnarray}
which tends to zero as $M(n)/D(n)=\omega(1)$. We remark that the
event $P_{I,(l)}^{k}>(1+\delta_5)E\left[P_{I,(l)}^{k}\right]$ for
all $l\in\{0,\cdots,\bar{l}D(n)\}$ is a sufficient condition for the
event $P_I^k\le(1+\delta_5)E[P_I^k]$. Thus, by the union bound over
all layers (including the cell with a desired transmitter), we have
the following inequality:
\begin{eqnarray}
\Pr\bigl\{\!\!\!\!\!\!\!\!\!&&P_{I,(l)}^k\le(1+\delta_5)E\left[P_{I,(l)}^k\right]\text{~~for
all $l\in\{0,\cdots,\bar{l}D(n)\}$}\bigr\} \nonumber\\ &&\ge 1-
\sum_{l=0}^{\bar{l}D(n)}\Pr\left\{P_{I,(l)}^k>(1+\delta_5)E\left[P_{I,(l)}^k\right]\right\}
\nonumber\\ &&=1-
(1+\bar{l}D(n))(1+\delta_5)^{8c_{10}\frac{M(n)}{D(n)}}
\nonumber\\
&&\quad\cdot\exp\left(-8c_{10}\frac{M(n)}{D(n)}(\delta_5-\epsilon_1)\right),
\nonumber
\end{eqnarray}
where the first inequality holds since there exist $\bar{l}D(n)$
layers, for some $\bar{l}>0$ independent of $n$. Finally, using the
union bound over $n$ nodes in the network yields that the total
interference power $P_I^k$ at receiver node $k$ is given by
$O(E[P_I^k])$ ($=O(P(n)M(n)D(n)^{\alpha-2})$) with probability of at
least
\begin{equation}
1-2\bar{l}nD(n)(1+\delta_5)^{8c_{10}\frac{M(n)}{D(n)}}\exp\left(-8c_{10}\frac{M(n)}{D(n)}(\delta_5-\epsilon_1)\right),
\nonumber
\end{equation}
which tends to one as $nD(n)\delta_2^{-M(n)/D(n)}=o(1)$ for a
certain constant
\begin{equation}
1<\delta_2<\left(\frac{e^{\delta_5-\epsilon_1}}{1+\delta_5}\right)^{8c_{10}}.
\nonumber
\end{equation}
This completes the proof of this lemma.


\newpage


\end{document}